# Spin-orbit torque magnetization switching in MoTe$_2$/permalloy heterostructures

*Shiheng Liang[1,2]†, Shuyuan Shi[1]†, Chuang-Han Hsu[3,4], Kaiming Cai[1], Yi Wang[1], Pan He[1], Yang Wu[1], Vitor M. Pereira[3,4], and Hyunsoo Yang[1,4]\**

[1]Department of Electrical and Computer Engineering, National University of Singapore, 117576, Singapore

[2]Faculty of Physics and Electronic Science, Hubei University, Wuhan 430062, People's Republic of China

[3]Department of Physics, National University of Singapore, 117542, Singapore

[4]Centre for Advanced 2D Materials, National University of Singapore, 117546, Singapore

**The ability to switch magnetic elements by spin-orbit-induced torques has recently attracted much attention for a path towards high-performance, non-volatile memories with low power consumption. Realizing efficient spin-orbit-based switching requires harnessing both new materials and novel physics to obtain high charge-to-spin conversion efficiencies, thus making the choice of spin source crucial. Here we report the observation of spin-orbit torque switching in bilayers consisting of a semimetallic film of 1T'-MoTe$_2$ adjacent to permalloy. Deterministic switching is achieved without external magnetic fields at room temperature, and the switching occurs with currents one order of magnitude smaller than those typical in devices using the best-performing heavy metals. The thickness dependence can be understood if the interfacial spin-orbit contribution is considered in addition to the bulk spin Hall effect. Further threefold reduction in the switching current is demonstrated with resort to dumbbell-shaped magnetic elements. These findings foretell exciting prospects of using MoTe$_2$ for low-power semimetal material based spin devices.**

Spintronics relies on both the electron's charge and spin degrees of freedom to develop functional electronic devices for the future needs in information and communication technologies. A crucial milestone in this direction is the manipulation of magnetic moments by planar electrical currents[1-3]. The simplest heterostructures interface a ferromagnetic strip with a conducting, non-magnetic film such that the spin Hall effect[4-6] and/or Rashba-Edelstein effect[6-8] induce a torque in the adjacent magnet. This has spurred magnetic memory concepts based on planar heterostructures[1, 9-11]. Although different microscopic mechanisms contribute to the magnetic torque[10], the central ingredient is the development of a robust non-equilibrium spin texture and accumulation at the interface between the conducting and recording layers in response to a charge current. This relies on the spin-orbit coupling (SOC) of conduction electrons in the non-magnetic layer, and the non-magnet is thus the critical component: not only should it feature a high charge-to-spin conversion efficiency, it must desirably have a low resistivity as well to enable low-power operation[3].

Spin-orbit torques (SOT) due primarily to SOC have recently become a topic of intense focus[10, 11]: On the one hand, torques arising from the spin-Hall (SH) effect in adjacent heavy metals such as Ta, Pt and W[1, 2, 12-14] appear to require large switching currents; this drives an impetus to discover new materials capable of overcoming this limitation. On the other hand, a number of emerging two-dimensional (2D) crystals and topological materials are promising platforms for spintronics[15, 16] due to their large intrinsic SOC and nontrivial spin textures[17-19]. In addition, crystal symmetries can be explored to engineer charge-to-spin conversion and SOT in device configurations that have been so far inaccessible.

Beyond large intrinsic SOC, some 2D materials display unique electronic structures, nontrivial spin textures, and spin-polarized surface states[20-22]. Most importantly, in the specific case of 2D semimetals, the combination of small conductivity and large SOC suggests that high charge-to-spin conversion efficiencies might be possible. Compared with topological insulators based on $Bi_2Se_3$[18], for example, semimetals display ten times smaller electrical resistivities, suggesting that SOT-based magnetization switching could operate with smaller power requirements. Previous studies of SOT based on 2D materials have been reported in semiconducting $MoS_2$ and $WSe_2$[23, 24], metallic $NbSe_2$[25], semimetallic $WTe_2$[26-28] and $MoTe_2$[29, 30]. Especially, the novel observation of out-of-plane spin induced anti-damping torque in $WTe_2$ and $MoTe_2$ hold a great promise for field-free switching of the perpendicularly magnetized system. However, a direct demonstration of SOT-induced magnetization switching utilizing a 2D material is still at its infancy and the thickness dependence of SOTs in 2D materials is not well understood.

Here, we employ the semimetal 1T'-$MoTe_2$ as a spin source and observe a large charge-to-spin conversion efficiency attributed to an underlying electronic structure that generates both a robust spin Hall conductivity (SHC) and interfacial spin accumulation. We report the direct observation of current-driven magnetization switching via SOT in a $MoTe_2$/Py heterostructure without magnetic fields at room temperature. The threshold current density required for complete magnetization reversal is found to be ~$10^5$ A/cm$^2$. First-principles calculations of SHC and interfacial spin accumulation in $MoTe_2$ provide a perspective over the type and symmetry of the torques which can account for our measurements. Finally, deliberately shaping the magnetic elements to confine their domain walls to the transport channel, we can decrease the critical switching current by a factor of 3.

We study devices consisting of $MoTe_2$ ($t$)/Py (6 nm) bilayers (Py = $Ni_{81}Fe_{19}$) with relatively thick $MoTe_2$ of $t$ = 65–110 nm in order to generate enough spin currents for magnetization switching with a reasonable resistivy to avoid the current shunting effect. The current flows along the $x$ direction (**Figure 1a**), the crystallographic $a$ axis of $MoTe_2$. The characterization of the $MoTe_2$ samples and fabrication of the devices are discussed in Supporting Information S1. Figure 1a shows the spin-torque ferromagnetic resonance (ST-FMR) measurement setup to estimate the SOT acting on the Py. A RF charge current $I_{RF}$ is injected into the bilayer and, if spin accumulation or spin currents develop at the $MoTe_2$/Py interface, spins from MoTe$_2$ diffuse into the Py layer and act upon its magnetic moment through an overall damping-like torque ($\tau_{DL}$) and/or a field-like torque ($\tau_{FL}$), both sinusoidal in time. The torque-induced precession of the magnetic moments causes an oscillatory modulation of the resistance of device which, when combined with the applied $I_{RF}$, generates a dc voltage, $V_{mix}$, the ST-FMR signal. An in-plane external magnetic field is swept over an angle $\theta$ with respect to the current direction $x$ to match the ferromagnetic resonance condition. A typical $V_{mix}$ signal shown in Figure 1b is analyzed by $V_{mix}(H) = V_S F_S + V_A F_A$, where $F_S$ and $F_A$ are symmetric and antisymmetric Lorentzian functions, respectively. The value of the effective spin Hall angle ($\xi_{SH}$) from our ST-FMR is 0.27 for the MoTe$_2$ (83.1 nm)/Py sample (Supporting Information S2), which is similar to the spin-to-charge conversion efficiency of 0.21 reported in the MoTe$_2$ stacked with graphene lateral spin valves.[30]

The fitting of the ST-FMR signal is also used to estimate the amplitudes of the in-plane and out-of-plane torques (cf. Figure 1a), whose dependence on the field orientation should be $\tau_\parallel = \tau_{DL}\cos(\theta)$ and $\tau_\perp = \tau_{FL}\cos(\theta) + \tau_\beta$, respectively[26]. We obtain a ratio $\tau_\parallel/\tau_\perp = 0.55$. The contribution $\tau_\beta$ is an out-of-plane anti-damping-like torque which has been previously detected in WTe$_2$/Py with currents applied along the low-symmetry $a$ axis[26] (Supporting Information S3). Since our current channels are defined along the low-symmetry $a$ axis of MoTe$_2$, we investigate the presence of such torque contribution in our MoTe$_2$/Py structures by fitting the angular dependence of the antisymmetric component to $V_A(\theta) = A\cos(\theta)\sin(2\theta) + B\sin(2\theta)$, as shown by the red curve in Figure 1c. In the thin region of $\beta$-MoTe$_2$ as reported by Stiehl *et al.*[29], there is an out-of-plane antidamping torque present in MoTe$_2$/Permalloy heterostructures, while in our case, we obtain $B \sim 0$, which indicates that $\tau_\beta$ could be negligible in a relative thick region of MoTe$_2$.

To characterize the magnetic anisotropy of the Py layer, the ST-FMR data have been analyzed for the dependence of the magnetic resonance frequency on the direction of in-plane magnetization, as shown in Figure 1d, with current along the *a* axis. Even though the shape anisotropy alone would determine the easy axis of Py to lie along the *x* or *a*-axis direction, the magnetic anisotropy induced by the proximal MoTe$_2$ results in the easy axis along *y*, parallel to the *b* crystallographic direction of MoTe$_2$. Direct measurements of the magneto-optical Kerr effect (MOKE) on our MoTe$_2$/Py devices further corroborate that the *b* axis of MoTe$_2$ is the easy axis in the Py layer (Supporting Information S4).

We have studied the magnetization switching response to pulsed dc currents (*I*) at room temperature. The magnetization of Py is collinear with the spin polarization in the semimetal, which implies that an anti-damping torque can switch the Py magnetization without external magnetic fields. The switching events have been captured by MOKE imaging in **Figure 2a-f**. The top row (Figure 2a-c) demonstrates that the magnetic moment of Py can be switched by pulsed current (30 μs pulse width) applied between the two electrodes outlined in green (the MoTe$_2$/Py bilayer is outlined in red). Prior to each set of measurements, the Py magnetization is saturated along the –*y* axis with an external magnetic field ($B_\text{ext}$), which is subsequently removed before applying the current pulses. Figure 2a shows the MOKE image after that initial step, under zero current, zero field, and saturated magnetization. Note that, in these MOKE images, dark regions correlate with magnetization along –*y* and grey regions indicate the magnetization along +*y*. Under these initial conditions and sending a current pulse along the –*x* direction, an area of reversed magnetization (grey) nucleates and gradually expands with increasing the current, as seen in Figure 2b-c. Figure 2c shows that the Py magnetization has undergone the complete reversal from –*y* (black block arrows) to +*y* (white block arrow) when the current density reached $J_\text{MoTe2} = 1.26 \times 10^6$ A/cm$^2$ ($J_\text{total} = 1.88 \times 10^6$ A/cm$^2$). Conversely, Figure 2d-f show the completion of an opposite magnetization switching cycle where, reversing the current to flow along the +*x* direction, flips the Py magnetization from +*y* back to –*y*.

The current-driven switching has been reproduced on a number of different devices with varying the MoTe$_2$ thickness (Supporting Information Figure S6). Figure 2g summarizes the critical switching current extracted for the bilayer ($J_{total}$, the totally current density is defined by $I/(w \cdot d)$, where $I$ is the current, $w$ is the width of device, and $d$ is the total thickness of the MoTe$_2$/Py bilayer) and for the semimetal film ($J_{MoTe2}$) as a function of MoTe$_2$ thickness. Both current densities increase with thickness. We note that the current density required for complete magnetization reversal is low, at $J_{MoTe2} = 6.71 \times 10^5$ A/cm$^2$ for a 66.1 nm thick MoTe$_2$ film. This value is one order of magnitude below that needed if the best heavy metals are used as spin sources[1, 2, 12]. For a comparison, we also perform current-driven switching measurements on control samples of Pt/Py and Cu/Py (Supporting Information S7). The magnetization switching takes place at $J_{total} \sim 2.8 \times 10^7$ A/cm$^2$ in the Pt (6 nm)/Py (6 nm) sample, which is about one order larger than the $J_C$ in MoTe$_2$/Py samples. We do not observe any switching behaviour in Cu (12 nm)/Py (6 nm) with a current density up to $J_{total} = 2.97 \times 10^7$ A/cm$^2$, which is one order of magnitude larger than the $J_C$ for our MoTe$_2$/Py samples.

Figure 2h displays the estimated SOT efficiency, which we quantify as an effective $\xi_{SH}$ on the basis of the conventional anti-damping spin-torque-driven model of magnetization switching[1, 31] (Supporting Information S5). Consistently with the evolution of the switching currents, the SOT efficiency is smaller for thicker MoTe$_2$ films. We determine the SOT efficiency of our thinnest device (66.1 nm) to be $\xi_{SH} \sim 0.35$ which is larger than those of Ta ($\sim 0.15$)[1] or Pt ($\sim 0.06$)[12]. The extracted $\xi_{SH}$ from the switching measurements shows a good agreement with that from the ST-FMR data of $\xi_{SH} \sim 0.27$ for 83.1 nm MoTe$_2$. The decreasing trend in Figure 2h as the thickness increases is opposite to that seen in magnetization switching driven by an spin Hall effect (SHE) in the heavy metal layer and WTe$_2$[28], where the efficiency increases with thickness and saturates when $t$ exceeds the spin diffusion length, $\lambda_{sf}$[5]. These observations point to a possible influence of the interface between MoTe$_2$ and Py. This is supported by the fact that, if one considers only the contribution of the intrinsic SHE in MoTe$_2$, the theoretical value of the SH angle computed for bulk MoTe$_2$ ($\theta_{SH}$ = 0.13) is smaller than the experimentally estimated values, as indicated by the dashed line in Figure 2h. This theoretical value was obtained as the ratio $\theta_{SH} \equiv \left(\frac{2e}{\hbar}\right)\frac{J_{s,z}^y}{J} = \left(\frac{2e}{\hbar}\right)\frac{\sigma_{zx}^y}{\sigma^{exp}}$ between the calculated SHC ($\sigma_{zx}^y$) and the experimental charge conductivity ($\sigma^{exp}$). From experiments, we extract the SHC of ~250 ($\hbar/e$)$\Omega^{-1}$cm$^{-1}$, which is similar to the value of ~280 ($\hbar/e$)$\Omega^{-1}$cm$^{-1}$ reported by Safeer et al.[30] but larger than the previously reported value of ~29 ($\hbar/e$)$\Omega^{-1}$cm$^{-1}$ [29].

To understand the possible microscopic origin of the large experimental SOT efficiency, we calculate both the intrinsic SHC and non-equilibrium spin accumulation of MoTe$_2$ from first-principles (Supporting Information S9-S10). The crystal structure of monoclinic 1T'-MoTe$_2$ (space group P2$_1$/m, #11) is shown in **Figure 3a**, where the box highlights the three-dimensional unit cell that encompasses two inequivalent MoTe$_2$ layers (the commonly referred zigzag chains of the transition metal ion run along the *a* direction). Figure 3b shows a representative band structure computed for a 20-monolayer slab. The energy dispersion displays an underlying band inversion centered at the Γ point in the Brillouin zone. Both few-layer and bulk crystals are semimetals with a low carrier density that arises from small Fermi surface pockets centered along the ΓX line in reciprocal space, with holes (electrons) nearer (farther) to the Γ point.

Figure 3c summarizes our results of the SHC, where the SH tensor component $\sigma_{ij}^\alpha$ quantifies the generation of a spin current $J_{s,i}^\alpha$ that propagates along the Cartesian direction $u_i$ carrying spins polarized along $u_\alpha$, as a result of an electric field pointing along $u_j$: $J_{s,i}^\alpha = \sigma_{ij}^\alpha E_j$. The experimental measurement layout with charge flowing along the *x* direction restricts the analysis to the SH components $\sigma_{zx}^y$ [32]. The component $\sigma_{zx}^z$ is not relevant, since our measurements reveal $\tau_\beta = 0$. It shows the intrinsic SHC calculated in Kubo linear response (Supporting Information S9) for charge neutrality (i.e. for chemical potential $\mu$ = 0) and varying thickness up to our numerical limit of 40 monolayers (27.7 nm). While the variation is somewhat pronounced between 1–3 layers, it settles to an asymptotic value $\sigma_{zx}^y \sim$ 60 $(\hbar/e)\,\Omega^{-1}\mathrm{cm}^{-1}$ within a few nm. By repeating the calculation at different chemical potentials, we have confirmed that the sign of the SH response is robust both as a function of thickness and doping, and also that its magnitude can be increased (decreased) under electron (hole) doping (Supporting Information Figure S9). Even though the range of thicknesses covered in the calculations is thinner than that of our experimental devices, these theoretical results unveil that the SHC of MoTe$_2$, $\sigma_{zx}^y$ has a considerable magnitude.

The inversion symmetry of the 1T' crystalline structure is broken in our bilayer devices. Combined with the relatively large SOC that one can infer from recent experimental values of $\xi_{\mathrm{SH}}$ in MoTe$_2$,[29, 30] in addition to the SHE, we inquire a non-equilibrium spin accumulation arising from the Rashba-Edelstein effect[7]. Figure 3d shows the thickness dependence of spin accumulation collinear with the magnetization in Py, $\delta s_a^b$,[33] computed at charge neutrality as well as $\mu = \pm 60$ meV. For a direct comparison with SHC, we introduce a spin diffusion current as $J_{s,\mathrm{diff}}^\alpha \equiv D\delta S_i^\alpha/\lambda_{\mathrm{sd}} = \delta s_i^\alpha E_i$, where $D = \lambda_{\mathrm{sd}}^2/\tau_s$ is a spin diffusion constant, $\lambda_{\mathrm{sd}}$ and $\tau_s$ are the spin diffusion length and relaxation time, respectively. This allows us to present the results in terms of the quantity $\delta s_i^\alpha$, which has dimensions of a spin conductivity. We used $\tau_s = \tau$ and $\lambda_{\mathrm{sd}} = 10$ nm, but different values can be considered simply rescaling the curves shown by the desired prefactor $\lambda_{\mathrm{sd}}\tau/\tau_s$, where $\tau$ is the Drude relaxation time; the value of $\lambda_{\mathrm{sd}}$ was estimated combining a recent experiment[30] with our calculated value of $\theta_{\mathrm{SH}}$ in the bulk; the assumption $\tau_s \approx \tau$ is expected for Elliot-Yaffet relaxation with strong spin-orbit coupling[34]. As anticipated, the spin accumulation decays quickly with increasing the thickness (note that $\delta s_i^\alpha$ is normalized by the thickness), reflecting the surface nature of this contribution. More importantly, we can see that the magnitudes of the accumulation-derived spin conductivity, $\delta s_a^b$ shown in Figure 3d for the thinnest slabs (< 4 nm) match those of the SHC plotted in Figure 3c. It indicates that interfacial spin accumulation contributes, at least, on equal footing with the SHE to the torque, and might even be the dominant effect if disorder is found to significantly reduce the intrinsic SHC from the values shown in Figure 3c.

We now demonstrate that the threshold switching current can be further decreased by engineering the shape of the heterostructure. **Figure 4a** shows dumbbell-patterned heterostructures of MoTe$_2$ (109.9 nm)/Py (6 nm), where the diameters of the left (right) disks are 16.60 μm (15.60 μm) and the channel length (width) is 11.50 μm (9.11 μm). This geometry reduces the domain nucleation energy that must be overcome to initiate the magnetization switching process by localizing the domain wall within the central channel. If one begins from the fully saturated state, a current density of $J_{\text{MoTe2}} \simeq 1.32 \times 10^6 \text{ A/cm}^2$ is needed to fully reverse the initial Py magnetization (Supporting Information S6). Figure 4b-h show MOKE images of the SOT-driven domain wall motion, at zero magnetic field, driven by a pulsed dc current along the *a*-axis of MoTe$_2$. Beginning with the domain wall at the edge of the rightmost disk (Figure 4b), a leftward (–*x*) current displaces the domain wall to the left. Complete displacement of the wall to the leftmost edge occurs for $J_{\text{MoTe2}}$ between $4.78 \times 10^5 \text{ A/cm}^2$ and $5.45 \times 10^5 \text{ A/cm}^2$. This is a near threefold reduction in comparison with the value needed for complete switching from full saturation (no domain wall, $1.32 \times 10^6 \text{ A/cm}^2$). As we see in Figure 4e, the domain wall stops at the edge of the leftmost disk, which happens because of the domain wall surface energy and the smaller current density in the disks compared with the channel. Micromagnetic simulations[35] shown in Figure 4i,k and 4j,l confirm this behavior (details in Supporting Information S8) and show that the disk regions effectively oppose the penetration of the domain wall. Finally, reversing the current direction from the state in Figure 4e, one can displace the domain wall in the opposite direction until it stops at the edge of the rightmost disk (Figure 4f-h). Consequently, in this shape-engineered geometry, the magnetization of the central channel can be reversibly switched back and forth with even smaller current densities than those needed for rectangular strips with the same overall area (as those in Figure 2).

To conclude, we have for the first time demonstrated the current induced magnetization switching in MoTe2/Py bilayers at room temperature, in the absence of external fields. The associated charge-to-spin conversion efficiency is $\xi_{SH} \sim 0.35$ for our thinnest device. The threshold current for switching can be as low as $J_{\mathrm{MoTe2}} \simeq 6.71 \times 10^5$ A/cm² in a rectangular MoTe2 (66.1 nm)/Py (6 nm) heterostructure. Our calculations suggest that the thickness dependent phenomena can be understood if the interfacial Rashba-Edelstein contribution is considered in addition to the SHE. We have also reduced the threshold currents by one third by shaping the heterostructure as a dumbbell so as to constrain the motion of the domain walls during the reversal process. This work opens exciting prospects for employing MoTe2 as a platform for 2D semimetal-based spintronics.

**Supporting Information**
Supporting Information is available.

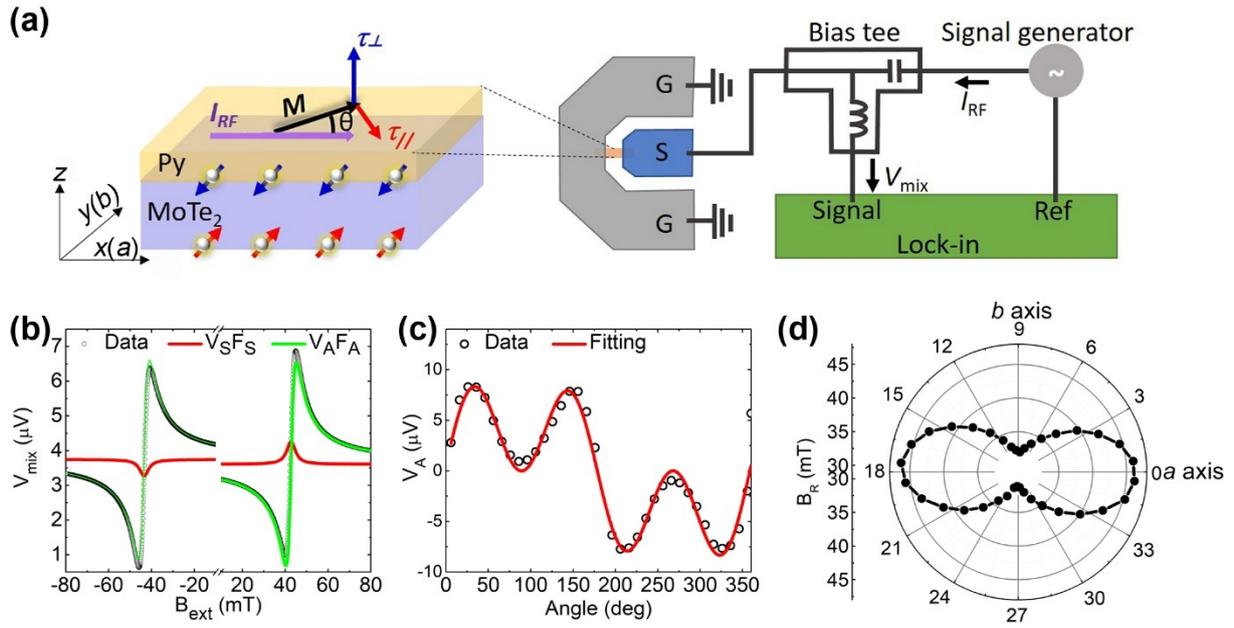

**Figure 1.** Spin torque ferromagnetic resonance measurements. a) Schematic of the ST-FMR measurement setup and illustration of the device with the SOT-induced magnetization dynamics. The directions *a* and *b* refer to the basal-plane crystal axes of monoclinic 1T'-MoTe$_2$ and are, respectively, oriented parallel to the *x* (longitudinal) and *y* (transverse) directions defined with respect to the current flow. b) A representative ST-FMR signal (open symbols) from a MoTe$_2$ (83.1 nm)/Py (6 nm) device with fits of the symmetric Lorentzian ($V_SF_S$) component (red lines), and the anti-symmetric Lorentzian ($V_AF_A$) component (green lines). The microwave frequency is 5 GHz and the applied microwave power is 14 dBm. An in-plane external magnetic field ($B_{ext}$) is applied at an angle $\theta = 36°$ with respect to $I_{RF}$. c) Amplitude of the anti-symmetric component of the resonance as a function of the angle of the in-plane field with the current applied parallel to the *a* axis. d) Ferromagnetic resonance field as a function of the in-plane magnetization angle, with the current applied parallel to the *a* axis. All measurements are done at room temperature.

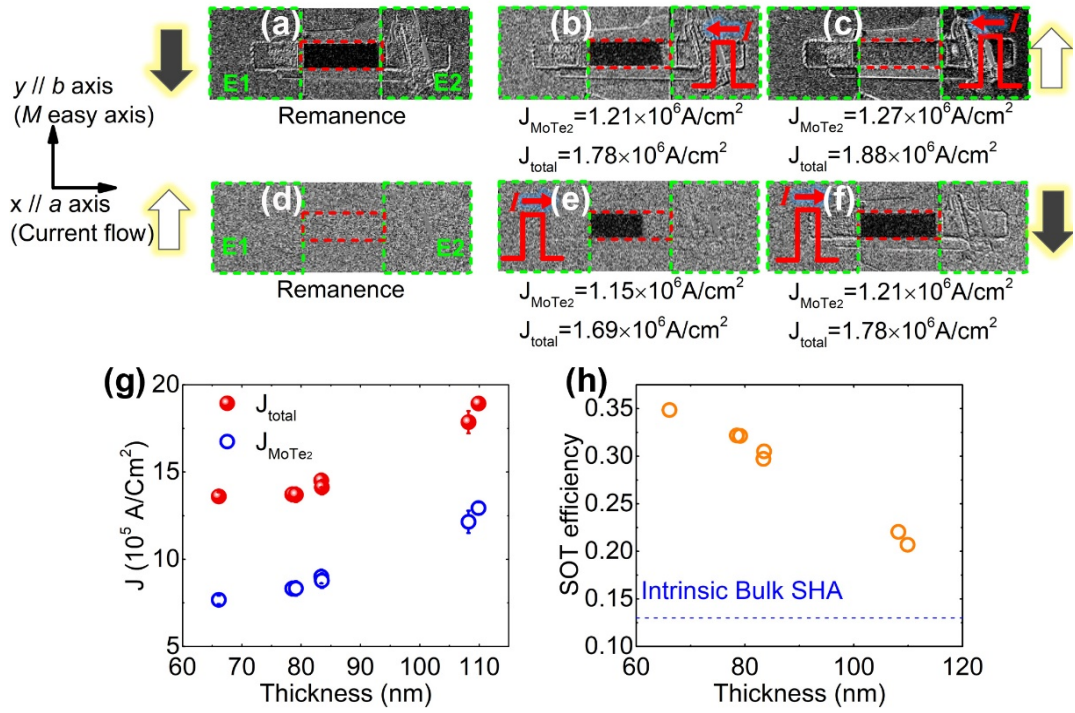

**Figure 2.** SOT-driven magnetization switching in MoTe$_2$ (108.2 nm)/Py heterostructures. a-c) MOKE images of magnetization switching (at zero magnetic field) under a pulsed dc current *I* along the –*x* direction (leftward in these figures; the magnetic easy axis of Py is along the *b* axis). Current densities increase from a) to c) and, when not zero, their values in the bilayer ($J_{total}$) and those estimated for the MoTe$_2$ layer alone ($J_{MoTe2}$) are stated underneath each panel. The red/dashed rectangles outline the 12 μm wide channel and the green/dashed ones mark the electrodes. The dark (light) contrast signals the local magnetization along −*y* (+*y*). White and black block-arrows are drawn to indicate the magnetization direction and its reversal in the panels with fully-switched states. d-f) MOKE images of the reverse switching process, where the initially magnetization state along +*y* is reversed by a current along +*x*. g) Thickness dependence of the switching current densities $J_{total}$ and $J_{MoTe2}$. h) Thickness dependence of the experimental SOT efficiency and the intrinsic bulk spin Hall angle (SHA) of MoTe$_2$ slabs calculated at the Fermi level. All measurements are done at room temperature.

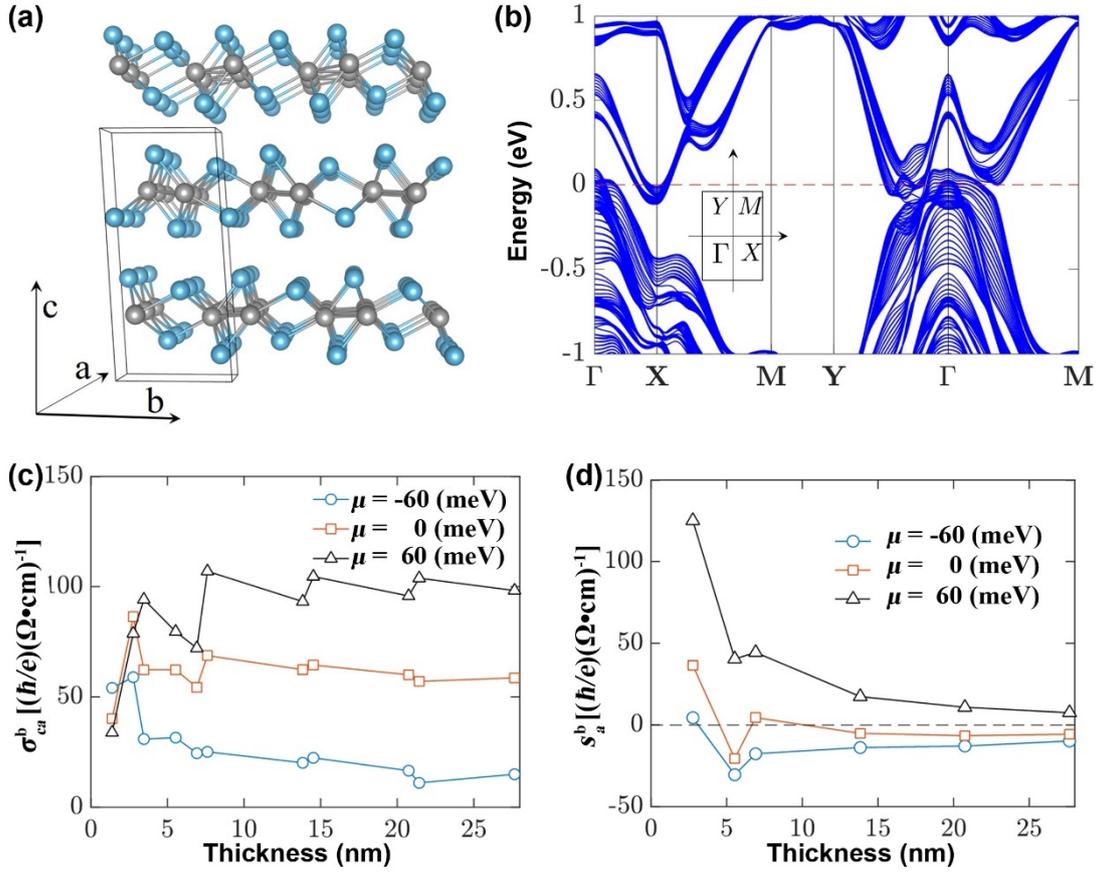

**Figure 3.** Electronic structure and spin response of MoTe$_2$. a) Crystal structure of the monoclinic 1T' phase of MoTe$_2$. b) Band structure of a MoTe$_2$ slab with 20 monolayers. c) The component $\sigma_{ca}^b$ ($= \sigma_{zx}^y$) of the SHC calculated for different thicknesses, at three different chemical potentials $\mu$ ($\mu = 0$ in undoped MoTe$_2$). d) The non-equilibrium spin density $\delta S_a^b$, which results from the accumulation of a finite spin density pointing along the $b$ crystalline direction under a charge current parallel to $a$ using $\tau_s = \tau$ and $\lambda_{sd} = 10$ nm.

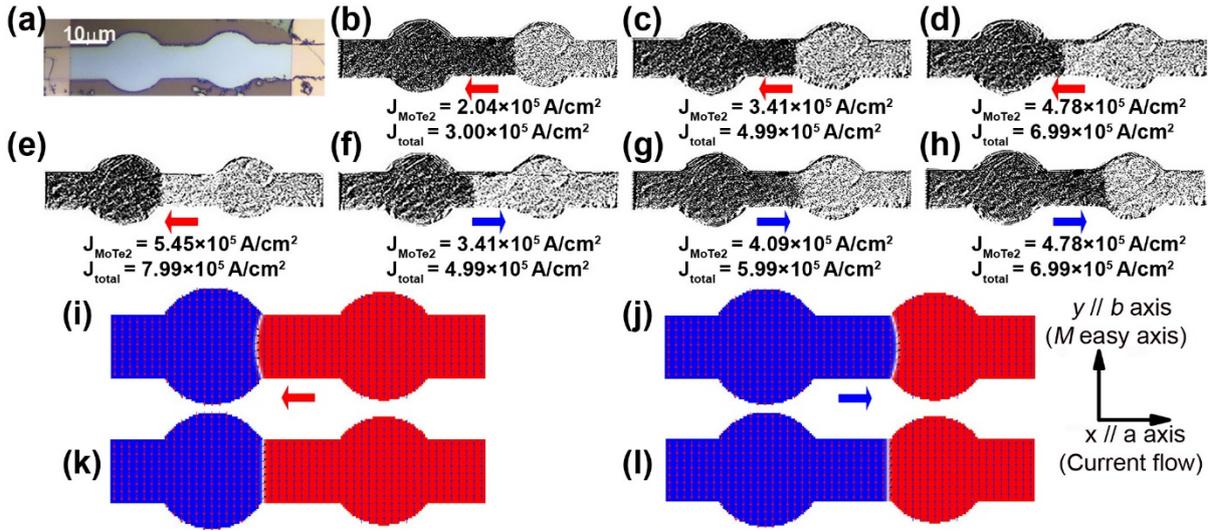

**Figure 4.** SOT-driven domain wall motion in dumbbell-patterned heterostructures of MoTe$_2$ (109.9 nm)/Py. a) Optical image of the dumbbell-shaped device. b-h) MOKE images of domain wall motion (at zero magnetic field) under a pulsed dc current *I* along the –*x* (b to e, red arrows) or +*x* directions (f to h, blue arrows) of MoTe$_2$. The current densities in the bilayer ($J_{total}$) and in the MoTe$_2$ layer ($J_{MoTe2}$) are stated underneath each panel. Arrows indicate the current direction. i-l) Micromagnetic simulation results of the SOT-driven domain wall motion. i) and j) show the domain wall being blocked at, respectively, the left and right edge of the disks under an applied dc current. k) and l) show the relaxed stable states after removal of the current.